\documentclass[aps,prx,twocolumn,showpacs,superscriptaddress,amssymb]{revtex4-1}

\usepackage{graphicx}
\usepackage{amsmath}
\usepackage{bm}
\usepackage{color}

\begin{document}

\title{Active Janus particles in a complex plasma}

\author{V. Nosenko}
\email{V.Nosenko@dlr.de}
\affiliation{Institut f\"{u}r Materialphysik im Weltraum, Deutsches Zentrum f\"{u}r Luft- und Raumfahrt (DLR), D-82234 We{\ss}ling, Germany}

\author{F. Luoni}
\email{f.luoni@gsi.de}
\affiliation{Institut f\"{u}r Materialphysik im Weltraum, Deutsches Zentrum f\"{u}r Luft- und Raumfahrt (DLR), D-82234 We{\ss}ling, Germany}
\affiliation{GSI Helmholtzzentrum f\"{u}r Schwerionenforschung, D-64291 Darmstadt, Germany}
\affiliation{Technische Universit\"{a}t Darmstadt, D-64277 Darmstadt, Germany}

\author{A. Kaouk}
\affiliation{Institut f\"{u}r Materialphysik im Weltraum, Deutsches Zentrum f\"{u}r Luft- und Raumfahrt (DLR), D-51170 Cologne, Germany}

\author{M. Rubin-Zuzic}
\affiliation{Institut f\"{u}r Materialphysik im Weltraum, Deutsches Zentrum f\"{u}r Luft- und Raumfahrt (DLR), D-82234 We{\ss}ling, Germany}

\author{H. Thomas}
\affiliation{Institut f\"{u}r Materialphysik im Weltraum, Deutsches Zentrum f\"{u}r Luft- und Raumfahrt (DLR), D-82234 We{\ss}ling, Germany}

\date{\today}
\begin{abstract}
Active Janus particles suspended in a plasma were studied experimentally. The Janus particles were micron-size plastic microspheres, one half of which was coated with a thin layer of platinum. They were suspended in the plasma sheath of a radio-frequency discharge in argon at low pressure. The Janus particles moved in characteristic looped trajectories suggesting a combination of spinning and circling motion; their interactions led to the emergence of rich dynamics characterized by non-Maxwellian velocity distribution. The particle propulsion mechanism is discussed, the force driving the particle motion is identified as photophoretic force.
\end{abstract}


\maketitle

\section{Introduction}

Active matter is a collection of active particles, each of which can convert the energy coming from their environment into directed motion, therefore driving the whole system far from equilibrium \cite{Elgeti:2015,Bechinger:2016}. Active matter has some intriguing physical properties and potentially a number of applications. It has recently become a hot topic of multiple interdisciplinary studies.

Complex plasmas, which are suspensions of micron-size solid particles in plasmas, are a particular instance of soft condensed matter \cite{Ivlev_book}. Complex plasmas are excellent model systems which are used to study various plasma-specific and generic phenomena at the level of individual particles. Their advantages include the possibility of directly observing virtually undamped dynamics of the particles suspended in a rarefied gas, in real time and with relative ease.

A particle in a complex plasma can become active via several mechanisms. First, the plasma wake effect, which makes the interparticle interaction nonreciprocal \cite{Melzer:1996,Lampe:2000,Ivlev:2015}, can under certain conditions lead to the particle self-propulsion. Examples include channeling particles \cite{Du:2014} and ``torsions'' \cite{Nosenko:2015}. Second, a particle can be driven by a phoretic force, e.g., the photophoretic force from the illumination laser \cite{Du:2017,Wieben:2018}. Third, as a rather extreme case, a particle can be propelled by the ``rocket force'' due to the ablation and removal of the particle material by a powerful laser irradiation \cite{Krasheninnikov:2010,Nosenko:2010}.

A prominent example of particles that can be active in various environments is the so-called Janus particles (JP), which have two sides with different properties \cite{Walther:2013}. Janus particles where two sides were made of (or coated with) different metals were found to be active in certain aqueous solutions \cite{Bechinger:2016}. In this paper, we experimentally study Janus particles - polymer microspheres half-coated with a thin layer of platinum - suspended in a gas discharge plasma. We show that they become active in this environment and discuss the mechanism involved.

\section{Experimental method}

Our experimental setup was a modified Gaseous Electronics Conference (GEC) radio-frequency (rf) reference cell \cite{Nosenko:2010}. Plasma was produced by a rf capacitively coupled discharge at $13.56$~MHz in argon. The gas pressure was varied in the range of $p_{\rm Ar}=0.66$--$13.3$~Pa, the rf discharge power was in the range of $P_{\rm rf}=1$--$20$~W.

A manual particle dispenser mounted in the upper flange was used to inject dust particles into the plasma. The particles were suspended in the plasma sheath above the rf electrode. They were illuminated by a horizontal laser sheet with the wavelength $\lambda=660$~nm and maximum output power of $100$~mW.
The particles were imaged from above using the Photron FASTCAM mini WX100 camera paired with the long-distance microscope Questar QM100 (or the Nikon Micro-Nikkor $105$-mm lens fitted with a matched bandpass interference filter in the experiment with a 2D layer of particles).
The particle coordinates and velocities were calculated in each frame with subpixel resolution using a moment method \cite{SPIT}. Note that the present optical setup does not allow us to resolve the particle spin.

Janus particles were prepared using the following method. Melamine-formaldehyde (MF) microspheres \cite{microparticles} with a diameter of $9.27~\mu$m and mass $m=6.3\times10^{-13}$~kg (respectively, $9.19~\mu$m and $6.14\times10^{-13}$~kg in the experiment with a 2D layer of particles) were dispersed in isopropanol. A drop of the mixture was placed on a Si wafer. After isopropanol evaporated, the particles formed a monolayer on the wafer surface. They were sputter-coated on one side with a $\approx 10$~nm layer of platinum. The resulting Janus particles were then collected by scratching them off the wafer by a sharp blade, placed in a standard container and dispensed into the plasma in a regular way.

\section{Results}

Two-dimensional (2D) suspensions of regular MF microspheres in plasmas are well studied both experimentally and theoretically \cite{Ivlev_book}. Therefore, as a first test with Janus particles we studied their 2D suspension. The experimental procedure was similar to that of Ref.~\cite{Couedel:2010}. A 2D layer of $9.19~\mu$m Janus particles was suspended in argon plasma at the low pressure of $p_{\rm Ar}=0.66$~Pa and discharge power of $P_{\rm rf}=20$~W. Top-view video of the particle suspension was recorded at the rate of $250$ frames per second.

\begin{figure}[tb]
\centering
\includegraphics[width=0.9\columnwidth]{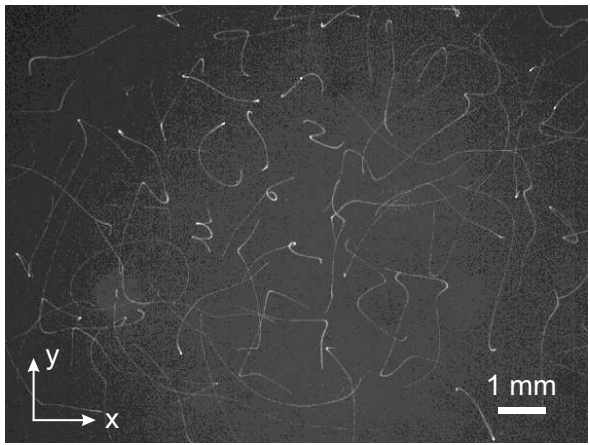}
\caption {\label {fig1} Trajectories of Janus particles suspended as a 2D layer in rf plasma sheath. $545$ frames of the top-view video (during $2.18$~s) were superposed, the brightness and contrast were adjusted for better viewing. Note a characteristic curly appearance of many trajectories. The argon pressure was $p_{\rm Ar}=0.66$~Pa, the rf discharge power was $P_{\rm rf}=20$~W.
}
\end{figure}

We observed that the particles energetically moved around in unusual curly trajectories, see Fig.~\ref{fig1}. Such trajectories are not normally seen for regular MF particles in similar experiments. This is a clear hint that JPs suspended in plasma behave as {\it circle swimmers} - a kind of active particles which tend to perform circular motion \cite{Jahanshahi:2017}. This hypothesis needs further study.

\begin{figure}[tb]
\centering
\includegraphics[width=0.9\columnwidth]{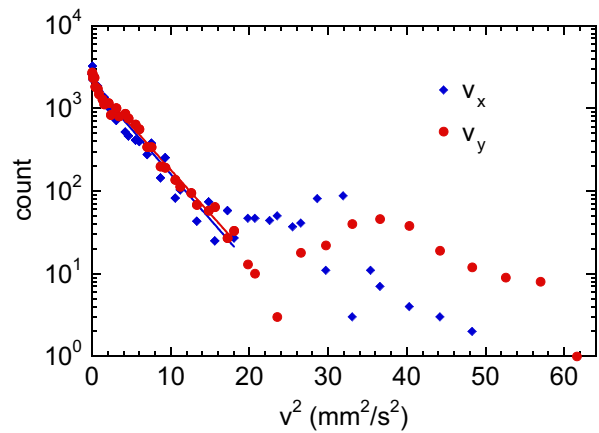}
\caption {\label {fig2} Distributions of the particle velocity $v_{x,y}$ in the experiment of Fig.~\ref{fig1}. The lines are Gaussian fits for $v<4.2$~mm/s. The apparent anisotropy of the particle velocity is probably due to their anisotropic illumination by the laser which shined in the $y$ direction.
}
\end{figure}

The distributions of the particle velocity $v_{x,y}$ are shown in Fig.~\ref{fig2}. Up to $v\approx4.2$~mm/s, they are on average Maxwellian with corresponding effective temperatures of $T_x=7.6$~eV and $T_y=7.9$~eV, which are much higher than that of the background gas. For $v>4.2$~mm/s, the distributions deviate substantially from the Maxwellian form and develop an extra maximum at $5.5$--$6$~mm/s. This is different from the regular MF particles in similar experiments, see e.g. Ref.~\cite{Nosenko:2006}. These observations suggest that there must be some kind of energy input or external drive on the system of Janus particles. To rule out unstable discharge conditions (e.g., caused by different secondary electron yields of aluminium and its oxide), an experiment with all-stainless-steel rf electrode system may be helpful. This is reserved for future work.

To get insight into the mechanisms responsible for the observed particle behaviour, we studied the most basic system, a single JP suspended in plasma \cite{FL_thesis}. This allowed us to exclude interparticle collisions and collective effects, e.g. instabilities mediated by the plasma wakes \cite{Couedel:2010}. Experimental parameters (argon pressure, illumination laser power) were varied in wide ranges and their effect on the Janus particles' behaviour was studied. Single JPs were trapped in a potential well created by a confining ring placed in the center of the rf electrode \cite{footnote0} and imaged by the top-view video camera operating at the rate of $60$ frames per second (see Ref.~\cite{Rubin-Zuzic:2018} for details of the experimental procedure).

\begin{figure}[b]
\centering
\includegraphics[width=0.9\columnwidth]{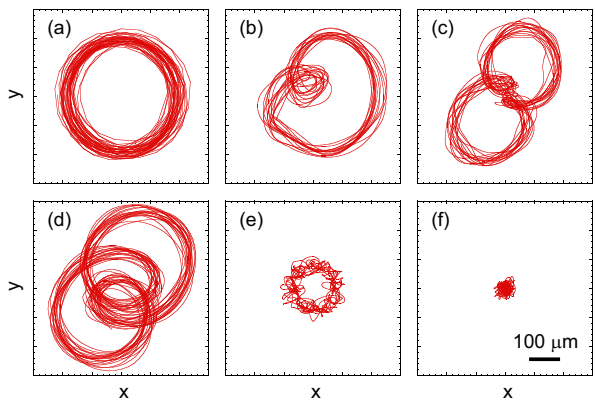}
\caption {\label {fig3} Typical trajectories of single Janus particles suspended in rf plasma sheath: (a) circular, (b-e) {\it epitrochoidal}, (f) apparently random. Regular MF particles have either random or sometimes ``smeared'' circular trajectories, as in panels (f) and (e), respectively.
}
\end{figure}

We observed that every single JP, when suspended in plasma, moved along a trajectory of one of the following three types:
(1) circular, (2) complex trajectories which are best described as {\it epitrochoids}, and (3) apparently random, see Fig.~\ref{fig3}. Out of the $21$ particles used in our experiments, $9$ had circular trajectories, $8$ epitrochoidal, and $4$ random trajectories.
The particle trajectory size, circling frequency, and other characteristics depended on the experimental conditions. The circling direction appeared randomly distributed between clockwise and counterclockwise.

For comparison, we performed similar experiments with single regular MF particles suspended in plasma. They had mostly random [as in Fig.~\ref{fig3}(f)] and sometimes ``smeared'' circular [as in Fig.~\ref{fig3}(e)] trajectories \cite{footnote1,Liu:2003}, but never clear epitrochoidal trajectories. The circling direction was either clockwise or counterclockwise. Furthermore, after properly centering the confining ring on the rf electrode, their trajectories were always random \cite{footnote2,Filliger:2007}. In contrast, JPs had circular or epitrochoidal trajectories with the same occurrence regardless of the exact centering of the confining ring.

The force driving a Janus particle can be either plasma-related (e.g., asymmetric ion drag) or caused by the illumination laser light (radiation pressure or photophoretic force). To discriminate between these forces, the experiment with a single JP was repeated using various combinations of the experimental parameters. The particle trajectories were recorded and from these, the average trajectory radius $r_{\rm tr}$ and circling frequency $\omega$ were calculated. They were typically in the ranges of $r_{\rm tr}=0.1$--$1.2$~mm and $\omega=1.2$--$17~{\rm s}^{-1}$. Note that for a uniformly circling particle, the tangential component of the driving force is given by $F_{\rm t}=m\gamma_{\rm E} \omega r_{\rm tr}$ (assuming balance of the driving force and gas friction). The Epstein neutral gas damping rate $\gamma_{\rm E}$ was calculated as in Ref.~\cite{Liu:2003}. We used $F_{\rm t}$ as a convenient measure of the driving force acting on a particle in the following experiments.

\begin{figure}[tb]
\centering
\includegraphics[width=0.9\columnwidth]{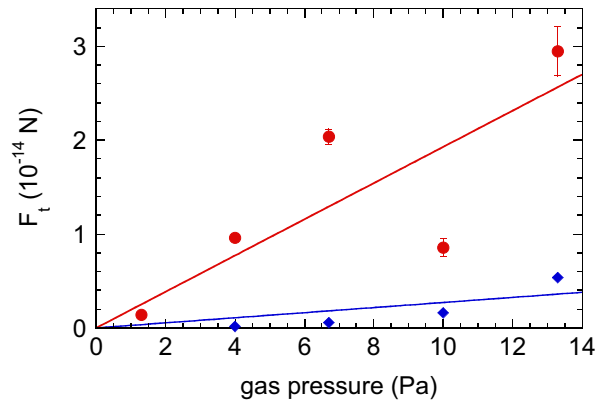}
\caption {\label {fig4} Tangential component $F_{\rm t}$ of the driving force acting on single Janus particles with circular trajectories as a function of the argon pressure $p_{\rm Ar}$. The particles were illuminated either by a laser with output power of $99$~mW (circles) or by a flashlight with a low-power incandescent lamp (diamonds). The lines are linear fits $F_{\rm t}\propto p_{Ar}$. The rf discharge power was $P_{\rm rf}=5$~W.
}
\end{figure}

For the particles with circular trajectories, $F_{\rm t}$ applies directly as a measure of the driving force. We performed a series of experiments with single JPs that had circular trajectories. The argon pressure was varied while keeping constant the rf discharge power.  The tangential component of the driving force $F_{\rm t}$ is shown in Fig.~\ref{fig4} as a function of the argon pressure, for two settings of the illumination laser power. (When the laser power was set to $0$, the particles were illuminated by a flashlight with a low-power incandescent lamp supplied with the microscope.) There is a clear trend for $F_{\rm t}$ to increase for higher gas pressures and higher illumination laser power in the ranges studied (except for one data point at $p_{\rm Ar}=10$~Pa).

\begin{figure}[tb]
\centering
\includegraphics[width=0.9\columnwidth]{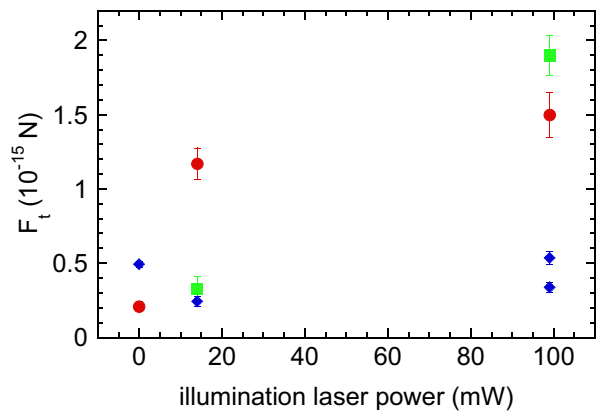}
\caption {\label {fig5} Tangential component $F_{\rm t}$ of the driving force acting on single Janus particles with epitrochoidal trajectories as a function of the illumination laser power. Data for three different particles are shown by different symbols. The argon pressure was $p_{\rm Ar}=1.3$~Pa, the rf discharge power was $P_{\rm rf}=5$~W.
}
\end{figure}

For the particles with epitrochoidal trajectories, $F_{\rm t}$ provides only an estimate of the driving force. For three different JPs with epitrochoidal trajectories, we measured $F_{\rm t}$ as a function of the illumination laser power at constant argon pressure and the rf discharge power, see Fig.~\ref{fig5}. While the data points scatter is large, there is a trend for the driving force to increase with the laser power for two out of three particles.

We observed that epitrochoidal trajectories become larger and cleaner as the illumination laser power is increased. When the laser is switched off, the trajectories always become circular.

These observations indicate that the illumination laser radiation plays an important role in the particle drive, at least for the particles with circular and epitrochoidal trajectories. The particles with random trajectories, on the contrary, did not show any remarkable trend with respect to the illumination laser power.

\begin{figure}[tb]
\centering
\includegraphics[width=0.9\columnwidth]{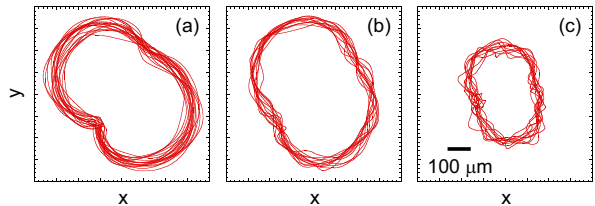}
\caption {\label {fig6} Trajectories of a single Janus particle in the experiment with varying discharge power: (a) $P_{\rm rf}=1$~W, (b) $5$~W, (f) $20$~W. The argon pressure was $p_{\rm Ar}=1.3$~Pa and the illumination laser power was $99$~mW.}
\end{figure}

In a separate test, the driving force on a single Janus particle with epitrochoidal trajectory diminished when the discharge power was increased in the range of $P_{\rm rf}=1$--$20$~W while keeping constant the argon pressure of $p_{\rm Ar}=1.3$~Pa and the illumination laser power of $99$~mW. This happened due to the decrease of both the trajectory size and circling frequency. The respective particle trajectories are shown in Fig.~\ref{fig6}. Since increasing $P_{\rm rf}$ results in a larger plasma density and therefore larger plasma-related forces such as the asymmetric ion drag force, these forces must be directed opposite to the main self-propulsion force, thus reducing the net force for larger $P_{\rm rf}$. In the experimental runs shown in Figs.~\ref{fig4} and \ref{fig5}, we used a middle setting of the discharge power of $5$~W.

\section{Particle propulsion mechanism}

The loops observed in the {\it epitrochoidal} particle trajectories, see Figs.~\ref{fig1} and \ref{fig3}(b),(c),(d), are a strong indication that these particles spin on top of their circling motion. It is natural to assume that the spinning frequency $\nu$ is equal to the circling frequency $\omega$ for circular trajectories, or is a multiple of the circling frequency ($\nu=n\omega$) for {\it epitrochoidal} trajectories, e.g., $n=2$ in Fig.~\ref{fig3}(b) and $n=3$ in Figs.~\ref{fig3}(c),(d). This kind of the coupled spinning and circling motion of a single driven particle (not of Janus type) was discussed in Refs.~\cite{Eymeren:2012,Jahanshahi:2017}. To drive this kind of particle motion, the spinning and circling frequencies must be locked and the spinning direction fixed and normal to the confining plane. In the simulations of Ref.~\cite{Jahanshahi:2017}, these conditions were set manually. For a Janus particle with one half of its surface coated, the spherical symmetry is broken and the axis connecting the centers of its two hemispheres is the axis of the self-propulsion force. This force is fixed in the particle frame of reference (the so-called body-fixed force). A Janus particle confined in the plasma potential trap will perform a circular motion, if the axis of the self-propulsion force is in the horizontal plane and directed tangentially to the particle's trajectory. For the particle to move in an epitrochoidal trajectory, this axis must additionally precess causing the particle spinning with a certain frequency with respect to the frequency of the circular motion. Since a direct observation of the particle spin would require a more advanced optical setup than used in the present paper, we leave the experimental verification and clarifying the physical origin of these important conditions for future work.

While the self-propulsion force can possibly arise from the asymmetry of many properties of a Janus particle halves (different surface charge densities, secondary electron emission coefficients, and temperatures, under electron and ion bombardments or laser illumination), our experimental observations point at the photophoretic force as the main driving mechanism for JPs. Indeed, the observed dependence of the driving force on the illumination laser power and gas pressure and the body-fixed nature of the force are consistent with the photophoretic force but not with the radiation pressure force (which is independent of the gas pressure). Photophoretic force is known to drive particles in various circling trajectories, e.g. ``circular'' and ``complex'' photophoresis in Ref.~\cite{Horvath:2014}. On the other hand, plasma-related forces such as the asymmetric ion drag force are apparently directed opposite to the photophoretic force thus reducing the net self-propulsion force. In addition, a possible mechanism of rotation excited by a combination of the vertical component of Earth's magnetic field and radial electric fields at the edge of the rf electrode \cite{Carstensen:2009} is ruled out by the apparently random circling direction of Janus particles observed in our experiments.

The photophoretic force acts on a nonuniform object surrounded by neutral gas when their temperatures are not equal \cite{Mackowski:1989,Horvath:2014,Nosenko:2010,Eymeren:2012,Beresnev:2012,Du:2017,Trott:2011,Kupper:2014}. There are two distinct varieties of the photophoretic force: $\Delta T$ and $\Delta \alpha$ forces. $\Delta T$ force arises when a particle is heated nonuniformly. The Pt-coated side of our Janus particles is expected to have a higher temperature than the opposite side. Indeed, in the experiments of Ref.~\cite{Nosenko:2010}, MF particles with thin Pd coating absorbed much more laser radiation than regular MF particles. Enhanced radiation absorbtion by the Pt or Pd coating may be due to the excitation of a surface plasmon in it \cite{Langhammer:2006}. $\Delta \alpha$ force arises when the thermal accommodation coefficient of the particle surface is not uniform. $\Delta \alpha$ photophoretic force, also called the ``accommodation force'' \cite{Beresnev:2012}, is a body-fixed force, which can provide propulsion and torque on the particles of irregular shape \cite{Eymeren:2012}. For Janus particles such as those used in our experiment, the $\Delta T$ photophoretic force can also be body-fixed.

The total photophoretic force (including the $\Delta T$ and $\Delta \alpha$ components) acting on a JP can be estimated using the following simple formula:

\begin{equation}
\label{eq:Fn}
\begin{split}
F_\textup{ph} & = \pi r_p^2 k_{\rm B} n_0 \Bigl(\alpha_1 T_1 + (1-\alpha_1) T_0 -\alpha_2 T_2 - (1-\alpha_2) T_0\Bigr)\\
 & = \pi r_p^2 k_{\rm B} n_0 T_0\Bigl(\alpha_1 \biggl(\frac{T_1}{T_0}-1\biggr) - \alpha_2\biggl(\frac{T_2}{T_0}-1\biggr)\Bigr),
\end{split}
\end{equation}

where $r_p$ is the Janus particle radius, $k_{\rm B}$ is Boltzmann's constant, $n_0$ and $T_0$ are respectively the neutral gas number density and temperature, $\alpha_{1,2}$ and $T_{1,2}$ are respectively the thermal accommodation coefficients and temperatures of the particle's two different halves. This formula assumes that $\alpha_{1,2}$ and $T_{1,2}$ are uniform over the particle halves and, in a major simplification, it does not answer the important question of how $T_{1,2}$ depend on the illumination intensity and the particle properties \cite{footnote3}, but it includes the essential physics of the photophoretic force \cite{footnote4} and can be used for estimates.

If a particle is heated uniformly, $T_1=T_2=T$, then $F_{\rm ph} = \pi r_p^2 k_{\rm B} n_0 (T-T_0)(\alpha_1-\alpha_2)$. This formula gives the pure $\Delta \alpha$ component of the photophoretic force. It vanishes, if the particle has the same temperature as the surrounding gas. If the accommodation coefficient is uniform over the particle surface, $\alpha_1=\alpha_2=\alpha$, then $F_{\rm ph} = \pi r_p^2 k_{\rm B} n_0 (T_1-T_2)\alpha$. This is the pure $\Delta T$ component. In real situations, both components will act simultaneously.

Reliable data on thermal accommodation coefficients even of common materials are hard to find in the literature. In a recent paper \cite{Trott:2011}, it was experimentally shown that for the range of materials and gases studied, the accommodation coefficient depends more on the gas type than on the surface material, implying that the role of the $\Delta \alpha$ force should be small. In Ref.~\cite{Beresnev:2012}, the $\Delta \alpha$ force (called ``accommodation force'' there) was experimentally estimated as $2$--$3$\% of the total photophoretic force acting on a steel particle in helium.

Assuming that the $\Delta T$ force is dominant in our experiments, we analyse the results shown in Fig.~\ref{fig4}. The lines here are fits $F_{\rm t}\propto p_{Ar}$, they fit the data points reasonably well. This kind of pressure dependence is consistent with the $\Delta T$ force. Using $\alpha=0.95$ for the accommodation coefficient of argon atoms on both Pt \cite{Trott:2011} and MF surfaces and assuming that argon behaves as an ideal gas at $T_0=300$~K, from the slope of the fit for the illumination laser power of $99$~mW we obtain $\Delta T\simeq10^{-2}$~K. This shows how sensitive the Janus particles are to a very small difference of temperature on their surface. For the case when the particle was illuminated by a low-power incandescent lamp, we calculate the temperature difference of $\Delta T\simeq10^{-3}$~K. This was probably due to the combined effect of illumination and plasma heating.

The photophoretic force and in particular the torque that it exerts on a particle strongly depend on small deviations from the surface homogeneity and spherical shape of the particle. This may explain the fairly large scatter of measurements in our experiments and also the existence of different types of trajectories for Janus particles from the same sample (the particles with random trajectories probably did not receive any coating at all during preparation due to their position on the wafer). This also makes the photophoretic force, especially of the $\Delta \alpha$ type, a strong candidate for the driving mechanism of the so-called ``abnormal'' particles, which are sometimes observed in complex plasma experiments \cite{Du:2017}. They move around with high speed and have involved trajectories. These particles may have inhomogeneities of their surface properties or deviations from the spherical shape \cite{Du:2017}, which give rise to the photophoretic force. Since Janus particles can be considered an extreme case of such inhomogeneous particles, they can be used as a study model for the ``abnormal'' particles, possibly helping to develop a method of controlling them in experiments with complex plasmas.

To summarize, we experimentally observed that Janus particles, which were plastic microspheres half-coated with a thin layer of platinum become active when suspended in argon discharge plasma. Single Janus particles perform periodic motion along circular or epitrochoidal trajectories. Photophoretic force from the illumination laser is proposed as the driving force providing a combination of propulsion and torque on the particles. In a 2D ensemble of Janus particles, their interactions lead to the emergence of rich dynamics characterized by non-trivial velocity distribution. Such ensembles of active Janus particles suspended in a plasma are promising model systems to study active Brownian motion \cite{Hagen:2011}, where the particle damping and propulsion can be tuned. An interesting topic of future research is building an analytical description of Janus particle collective dynamics \cite{Yazdi:2016}.

\section{Acknowledgments}

The authors thank S. Zhdanov, S. Khrapak, and S. Yurchenko for helpful discussions and M. Sperl for carefully reading the manuscript.


\begin{thebibliography}{}


\bibitem {Elgeti:2015} J. Elgeti, R. G. Winkler, and G. Gompper, Physics of microswimmers-single particle motion and collective behavior: a review, Rep. Prog. Phys. {\bf 78}, 056601 (2015).

\bibitem {Bechinger:2016} C. Bechinger, R. Di Leonardo, H. L\"{o}wen, Ch. Reichhardt, G. Volpe, G. Volpe, Active Particles in Complex and Crowded Environments, Rev. Mod. Phys. {\bf 88}, 045006 (2016).

\bibitem {Ivlev_book} A. Ivlev, H. L\"{o}wen, G. Morfill, C. P. Royall, {\it Complex Plasmas and Colloidal Dispersions: Particle-resolved Studies of Classical Liquids and Solids}, Series in Soft Condensed Matter Vol. 5 (World Scientific, Singapore, 2012).

\bibitem{Melzer:1996} A. Melzer, V. A. Schweigert, I. V. Schweigert, A. Homann, S. Peters, and A. Piel, Structure and stability of the plasma crystal, Phys. Rev. E {\bf 54}, R46 (1996).

\bibitem{Lampe:2000} M. Lampe, G. Joyce, G. Ganguli, and V. Gavrishchaka, Interactions between dust grains in a dusty plasma, Phys. Plasmas {\bf 7}, 3851 (2000).

\bibitem {Ivlev:2015} A. V. Ivlev, J. Bartnick, M. Heinen, C.-R. Du, V. Nosenko, and H. L\"{o}wen, Statistical Mechanics where Newton’s Third Law is Broken, Phys. Rev. X {\bf 5}, 011035 (2015).

\bibitem {Du:2014} C.-R. Du, V. Nosenko, S. Zhdanov, H. M. Thomas, and G. E. Morfill, Channeling of particles and associated anomalous transport in a two-dimensional complex plasma crystal, Phys. Rev. E {\bf 89}, 021101(R) (2014).

\bibitem {Nosenko:2015} V. Nosenko, S. K. Zhdanov, H. M. Thomas, J. Carmona-Reyes, and T. W. Hyde, Spontaneous formation and spin of particle pairs in a single-layer complex plasma crystal, EPL {\bf 112}, 45003 (2015).

\bibitem {Du:2017} C. R. Du, V. Nosenko, H. M. Thomas, A. M\"{u}ller, A. M. Lipaev, V. I. Molotkov, V. E. Fortov, and A. V. Ivlev, Photophoretic force on microparticles in complex plasmas, New J. Phys. {\bf 19}, 073015 (2017).

\bibitem {Wieben:2018} F. Wieben and D. Block, Photophoretic force measurement on microparticles in binary complex plasmas, Phys. Plasmas {\bf 25}, 123705 (2018).

\bibitem {Krasheninnikov:2010} S. I. Krasheninnikov, A. Yu. Pigarov, R. D. Smirnov, and T. K. Soboleva, Theoretical Aspects of Dust in Fusion Devices, Contrib. Plasma Phys. {\bf 50}, 410 (2010).

\bibitem {Nosenko:2010} V. Nosenko, A. V. Ivlev, and G. E. Morfill, Laser-induced rocket force on a microparticle in a complex (dusty) plasma, Phys. Plasmas {\bf 17}, 123705 (2010).

\bibitem {Walther:2013} A. Walther and A. H. E. M\"{u}ller, Janus Particles: Synthesis, Self-Assembly, Physical Properties, and Applications, Chem. Rev. {\bf 113}, 5194 (2013).

\bibitem {SPIT} U. Konopka ``Super Particle Identification and Tracking'' (unpublished).

\bibitem {microparticles} Marketed by Microparticles GmbH, see http://microparticles.de.

\bibitem{Couedel:2010} L. Cou\"edel, V. Nosenko, A. V. Ivlev, S. K. Zhdanov, H. M. Thomas, and G. E. Morfill, Direct Observation of Mode-Coupling Instability in Two-Dimensional Plasma Crystals, Phys. Rev. Lett. {\bf 104}, 195001 (2010).

\bibitem {Jahanshahi:2017} S. Jahanshahi, H. L\"{o}wen, and B. ten Hagen, Brownian motion of a circle swimmer in a harmonic trap, Phys. Rev. E {\bf 95}, 022606 (2017).

\bibitem {Nosenko:2006} V. Nosenko, J. Goree, and A. Piel, Laser method of heating monolayer dusty plasmas, Phys. Plasmas {\bf 13}, 032106 (2006).

\bibitem {FL_thesis} F. Luoni, Single Janus particles in a complex plasma environment, MS thesis, Politecnico di Milano, 2018.

\bibitem {footnote0} The confining ring was made of an aluminium alloy and had an internal diameter of $16$~mm and hight of $2$~mm.

\bibitem {Rubin-Zuzic:2018} M. Rubin-Zuzic, V. Nosenko, S. Zhdanov, A. Ivlev, H. Thomas, S. Khrapak, and L. Cou\"{e}del, Single Particle Dynamics in a Radio-Frequency Produced Plasma Sheath, AIP Conf. Proc. {\bf 1925}, 020023 (2018).

\bibitem {footnote1} ``Anomalous'' circling trajectories of single MF particles suspended in plasma and illuminated with laser light were previously reported in Ref.~\cite{Liu:2003}.

\bibitem {Liu:2003} B. Liu, J. Goree, V. Nosenko, and L. Boufendi, Radiation pressure and gas drag forces on a melamine-formaldehyde microsphere in a dusty plasma, Phys. Plasmas {\bf 10}, 9 (2003).

\bibitem {footnote2} One possible explanation of the particle circling in the shifted confining ring may be the so-called Brownian gyrator effect \cite{Filliger:2007} caused by the flow of ions at the particle location.

\bibitem {Filliger:2007} R. Filliger and P. Reimann, Brownian Gyrator: A Minimal Heat Engine on the Nanoscale, Phys. Rev. Lett. {\bf 99}, 230602 (2007).

\bibitem {Eymeren:2012} J. van Eymeren and G. Wurm, The implications of particle rotation on the effect of photophoresis, Mon. Not. R. Astron. Soc. {\bf 420}, 183 (2012).

\bibitem {Horvath:2014} H. Horvath, Photophoresis - a Forgotten Force ??, KONA Powder and Particle Journal {\bf 31}, 181 (2014).

\bibitem {Carstensen:2009} J. Carstensen, F. Greiner, L.-J. Hou, H. Maurer, and A. Piel, Phys. Plasmas {\bf 16}, 013702 (2009).

\bibitem {Mackowski:1989} D. W. Mackowski, Photophoresis of Aerosol-Particles in the Free Molecular and Slip-Flow Regimes, Int. J. Heat Mass Transfer, {\bf 32}, 843 (1989).

\bibitem {Beresnev:2012} S. Beresnev, M. Vasiljeva, D. Suetin, Predictions and detection of the ``accommodation'' forces on Janus particles subjected to directed radiation in a rarefied gas, Vacuum {\bf 86}, 1663 (2012).

\bibitem {Trott:2011} W. M. Trott, J. N. Casta\~{n}eda, J. R. Torczynski, M. A. Gallis, D. J. Rader, An experimental assembly for precise measurement of thermal accommodation coefficients, Rev. Sci. Instrum. {\bf 82}, 035120 (2011).

\bibitem {Kupper:2014} M. K\"{u}pper, C. de Beule, G. Wurm, L. S. Matthews, J. B. Kimery, T. W. Hyde, Photophoresis on polydisperse basalt microparticles under microgravity, Journal of Aerosol Science {\bf 76}, 126 (2014).

\bibitem{Langhammer:2006} Ch. Langhammer, Z. Yuan, I. Zori\'{c}, and B. Kasemo, Plasmonic properties of supported Pt and Pd nanostructures, Nano Letters {\bf 6}, 833 (2006).

\bibitem {footnote3} The particle properties are usually taken into account by introducing the asymmetry factor $J_1$, see, e.g. Ref.~\cite{Mackowski:1989}. Calculating $J_1$ for a given particle is not straightforward, however.

\bibitem {footnote4} This formula, however, does not capture the correct pressure dependence of the $\Delta \alpha$ force.

\bibitem {Hagen:2011} B. ten Hagen, S. van Teeffelen, and H. L\"{o}wen, Brownian motion of a self-propelled particle, J. Phys.: Condens. Matter {\bf 23}, 194119 (2011).

\bibitem {Yazdi:2016} A. Yazdi and M. Sperl, Glassy dynamics of Brownian particles with velocity-dependent friction, Phys. Rev. E {\bf 94}, 032602 (2016).

\end{thebibliography}
\end{document}